\documentclass{aa}  

\usepackage{graphicx}
\usepackage{txfonts}
\usepackage{natbib}
\usepackage{hyperref}
\usepackage{caption}
\usepackage{subcaption}
\usepackage{threeparttable}
\usepackage[switch, modulo]{lineno}

\begin{document} 

\title{The PLATO Simulator: modelling of high-precision high-cadence space-based imaging\thanks{Software package available at the {\sc PLATO Simulator} web site (\href{https://fys.kuleuven.be/ster/Software/PlatoSimulator/}{https://fys.kuleuven.be/ster/Software/PlatoSimulator/}).}}


   \author{P. Marcos-Arenal\inst{1}
           \and
          W. Zima\inst{1}
           \and
          J. De Ridder\inst{1}
           \and
          C. Aerts\inst{1,2}
           \and
          R. Huygen\inst{1}
           \and
          R. Samadi\inst{3}
           \and
          J. Green\inst{3}
           \and
          G. Piotto\inst{4}
           \and
          S. Salmon\inst{5}
           \and
          C. Catala\inst{3} 
          \and
          H. Rauer\inst{6,7}
         }

\institute{Instituut voor Sterrenkunde, KU Leuven, Celestijnenlaan 200D, 3001 Leuven, Belgium\\
              \email{Pablo.MarcosArenal@ster.kuleuven.be}
           \and
             Department of Astrophysics, IMAPP, Radboud University Nijmegen, 6500 GL Nijmegen, The Netherlands
           \and
             LESIA, Observatoire de Paris, CNRS UMR 8109, UPMC, Universit\'e Denis Diderot, 5 Place Jules Janssen, 92195 Meudon Cedex, France
           \and
             Dipartimento di Astronomia, Vicolo dell'Osservatorio 5, 35122 Padova, Italy
           \and
             Institut d'Astrophysique et de G\'eophysique de l'Universit\'e de Li\`ege, All\'ee du 6 Ao\^ut 17, 4000 Li\`ege, Belgium
           \and
            German Aerospace Center (DLR) Institut f\"{u}r Planetenforschung Extrasolare Planeten und Atmosph\"{a}ren, Rutherfordstra\ss e 2, 12489 Berlin, Germany
           \and
            TU Berlin, Hardenbergstr. 36, 10623 Berlin, Germany
               }

\date{Received ; Accepted}
   
\titlerunning{The PLATO Simulator}
\authorrunning{P.\ Marcos-Arenal et al.} 

 
\abstract
{Many aspects of the design trade-off of a space-based instrument and
its performance can best be tackled through simulations of the expected
observations. The complex interplay of various noise sources in the course of
the observations make such simulations an indispensable part of the assessment
and design study of any space-based mission.}  {We present a formalism to
model and simulate photometric time series of CCD images by including models
of the CCD and its electronics, the telescope optics, the stellar field, the
jitter movements of the spacecraft, and all of the important natural noise sources.}
{This formalism has been implemented in a versatile end-to-end simulation
  software tool, specifically designed for the PLATO (Planetary Transists and Oscillations of Stars) 
	space mission to be operated from L2, but easily adaptable to
  similar types of missions. We call this tool {\sc PLATO Simulator}.}  
 {We provide a detailed description of  several noise sources and discuss their properties in connection with the optical
design, the allowable level of jitter, the quantum efficiency of the detectors, etc. 
The expected overall noise budget of generated light curves is computed, 
as a function of the stellar magnitude, for different sets of input parameters describing the instrument properties. 
The simulator is offered to the scientific community for future use.}{}


\maketitle

%

\section{Introduction}

Recent uninterrupted long-term $\mu$-mag-precision space photometry opened
  a new era in time-domain astronomy and has led to numerous exoplanet detections,
  see, e.g., \citet{Moutou2013} for a review of CoRoT 
	(Convection, Rotation and planetary Transits) results and
  \citet{Borucki2010,Welsh2012,Batalha2013} for results obtained from the {\it
    Kepler\/} mission.  As a by-product, both space missions also implied a
  goldmine for stellar variability studies
  \citep[e.g.,][]{Debosscher2009,Sarro2009,Prsa2011,Debosscher2011}.  In
  particular, detailed seismic probing was finally  reached and gave new insights
  into the physics of stellar and galactic structure, pointing out limitations of
  standard models \citep[e.g.,][]{Degroote2010,Beck2012,Miglio2012,Miglio2013}.
  Even tests of stellar evolution theory for a wide variety of stellar masses
  and evolutionary stages, through asteroseismic data alone or combined with
  ground-based data, became possible thanks to dedicated CoRoT and {\it
    Kepler\/} asteroseismology programmes
  \citep[e.g.,][]{Michel2006,Gilliland2010,Bedding2011} and from multivariate
  statistical studies based on seismic, polarimetric, and spectroscopic data
  \citep[e.g.,][]{Aerts2014}.  Asteroseismology of eclipsing binaries
  \citep[e.g.,][]{Maceroni2009,Welsh2011,Tkachenko2014,Beck2014} and of
  exoplanet host stars
  \citep[e.g.,][]{Gilliland2011,Chaplin2013,Huber2013,HuberScience2013,VanEylen2014}
  only became possible in the current space photometry era.

Despite the availability of these numerous CoRoT and {\it Kepler\/} data sets
with long time-base, new projects for similar studies are in development. These new
studies are capable of mapping the entire sky rather than just a small portion of it, as was
the case for CoRoT and {\it Kepler}. The current paper
concerns the PLATO2.0 mission (hereafter simply called PLATO), which was
recently accepted as M3 mission in the Cosmic Vision 2015 -- 2025 programme of
the European Space Agency (ESA). PLATO is an acronym for PLanetary Transits and
Oscillations of Stars and is a mission that will operate from the second
Lagrange point (L2) of the Sun-Earth system.

PLATO's goals are to study the formation and evolution of planetary systems,
with specific emphasis on Earth-like planets in the habitable zone of bright
solar-like host stars. PLATO will have the capacity to
detect and characterize hundreds of  Earth-like planets and thousands of
larger planets with the photometric transit technique already used by CoRoT and
{\it Kepler}.  Up to 1\,000\,000 stars will be observed and characterized over
the course of the full mission.  Masses, radii, and ages of 80\,000 dwarf and
subgiant stars will be measured with sufficient precision to allow for their
asteroseismic modelling.  The expected noise level for stars with visual
magnitudes of less than 11 is 34 ppm per hour, and for stars brighter than 13th
magnitude the noise is expected to be below 80 ppm per hour. A unique feature of
PLATO compared to previous and other planned space missions is its capacity to
measure a fraction of the targets in two photometric bands.

In order to achieve its scientific aims, PLATO is equipped with 34 12\,cm
aperture telescopes and 136 CCDs (four CCDs per camera) with 4510$\times$4510 18
$\mu$m pixels, to cover about 50\% of the sky, operating in the 500-1000 nm
spectral range. Each selected target is assigned a 6$\times$6 pixel window to
produce its light curve on board. This on-board processing is required to limit
the amount of data to be downloaded to ground for its wide field of view (FoV).
Detailed descriptions of the PLATO M3 mission are available in \citet{Rauer2013}
and in the Yellow Book submitted to ESA for the selection of M3
\citep{Esa2013}\footnote{http://sci.esa.int/plato/53450-plato-yellow-book/}.

PLATO's cameras are high-precision imagers whose expected performance must be
carefully assessed from an appropriate overall instrument model.  The instrument
noise performance cannot be derived from the simple addition of the noise
properties of the individual contributors due to the complex interaction
between the various noise sources. As is often the case, it is not feasible to
build and test a prototype of the PLATO imaging devices. Hence, numerical
simulations performed by an end-to-end simulator are used to model the 
noise level expected to be present in the observations. Such simulations not
only allow us to study the performance of the instrument, its noise source response,
and the data quality, but they are also an essential tool for the fine-tuning of the
instrument design for different types of configurations and observing
strategies. The simulator should also allow us to test the scientific feasibility
of an observing proposal. In this way, a complete description and assessment of
the expected objectives of the mission can be derived.  

In this paper, we present a formalism, termed {\sc PLATO Simulator}, to model
each of the noise sources affecting a space-based high-resolution imager and
the mutual interaction of these noise sources.  The performance of previous space photometers, such
as MOST \citep{Walker2005}, CoRoT \citep{Auvergne2009}, and {\it Kepler}
\citep{Koch2010, Caldwell2010} have been tested and evaluated using approaches
specifically designed for each of these missions alone, keeping in mind their
orbit (low-Earth in the case of MOST (Microvariablity and Oscillations of Stars) 
and CoRoT and Earth-trailing for {\it
  Kepler}).

Our aim is to provide the scientific community with a tool that is easily
adaptable for other high-precision photometric space missions, taking PLATO as
the case study to illustrate our simulator.  Our approach here is based on
previous work developed in this spirit for the MONS (Measuring Oscillations in Nearby Stars) 
and Eddington mission
projects, which never made it to implementation phase \citep[][hereafter
  termed DAK06]{DeRidder2006}. We have further developed and
  implemented this formalism in the {\sc PLATO Simulator} end-to-end simulation software-tool,
  which was specifically constructed for the PLATO assessment and Phase A/B1
  studies, but is easily adaptable to other missions.  

In the following section, we describe each of the noise sources and the
algorithms developed to model them, as well as their implementation and
interaction.  We introduce the {\sc PLATO Simulator} in Sect.\,\ref{sec:package}
and, finally, in Sect.\,\ref{sec:application}, we present applications of the
simulator to the study of white noise and jitter for the PLATO mission. These
results were used to predict the quality of its photometry to assess
the transit and stellar variability detection capability and to provide
essential feedback for the mission design.


\section{Imaging simulation}

A previous preliminary version of the simulator \citep{Zima2010} has now
  been completely rebuilt. It relies on new architecture, to make it
  adaptable to other missions as well.  The technical details and motivation of
  the new architecture were already described in \cite{Marcos2013}, to which we
  refer the interested reader, and are therefore ommitted here.

\begin{figure*}[!ht]
\centering
\includegraphics*[width=158mm, clip,angle=0]{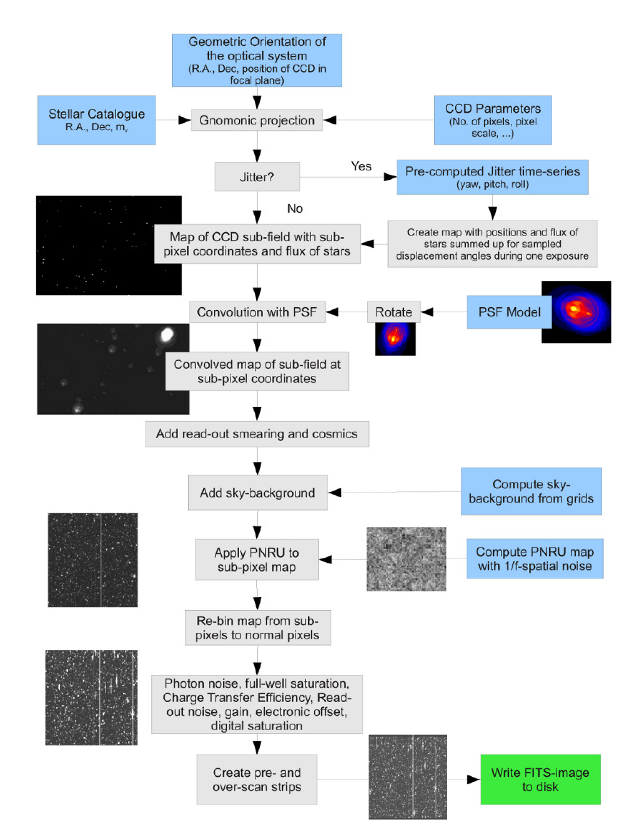}
\caption{Processing steps that are applied in the simulator to
  model a CCD image of a stellar field, in sequential
    order.}
\label{fig:flowdiagram}
\end{figure*}

The new {\sc PLATO Simulator} generates synthesized images by simulating the
acquisition process of a space-based detector instrument as realistically as
possible. Each image is numerically modelled, based on a number of input
parameters, which define the set-up of the CCD and its
electronics, the properties of the optical instrument, the FoV,
the point spread function (PSF), the pixel response non-uniformity (PRNU) and
all related noise sources.  The process of image generation can be classified
according to the sequential order shown in Fig.\,\ref{fig:flowdiagram}.
The following subsections will give a detailed description of each of these parameters:
\begin{itemize}
\item  Imaging FoV:
	\subitem CCD rotation and resizing;
	\subitem The CCD sub-pixel matrix;
	\subitem Mapping stars on the CCD;
	\subitem High-energy particle hits;
  \item  Satellite jitter;
  \item PSF Convolution;	
  \item  CCD Sensitivity variations: PRNU;
 \item   Noise effects:
	\subitem Read out smearing;
	\subitem Sky background;
	\subitem Photon noise;
	\subitem Electronic noise sources.
\end{itemize}
This subdivision is based on the processing chain of the whole
  simulation, whose schematic flow diagram is shown in
  Fig.\,\ref{fig:flowdiagram}.

\subsection{Imaging FoV}\label{sec:FoV}

A set of input parameters is required to generate a complete synthetic CCD
image. This set of parameters defines general properties of the telescope and
the focal plane, the orientation of the stellar field, and the properties of the
simulated time series, such as the exposure time and the number of exposures to
be computed. To model the CCD frame, the detector characteristics are taken as
inputs in terms of frame size, pixel size, and pixel scale, as well as the right
ascension and declination centre of the optical axis, the CCD orientation, the
focal plane coordinates, and the focal plane orientation.

\subsubsection{The CCD rotation and resizing}

\begin{figure}
\centering
\includegraphics[width=0.4\textwidth, clip, angle=0]{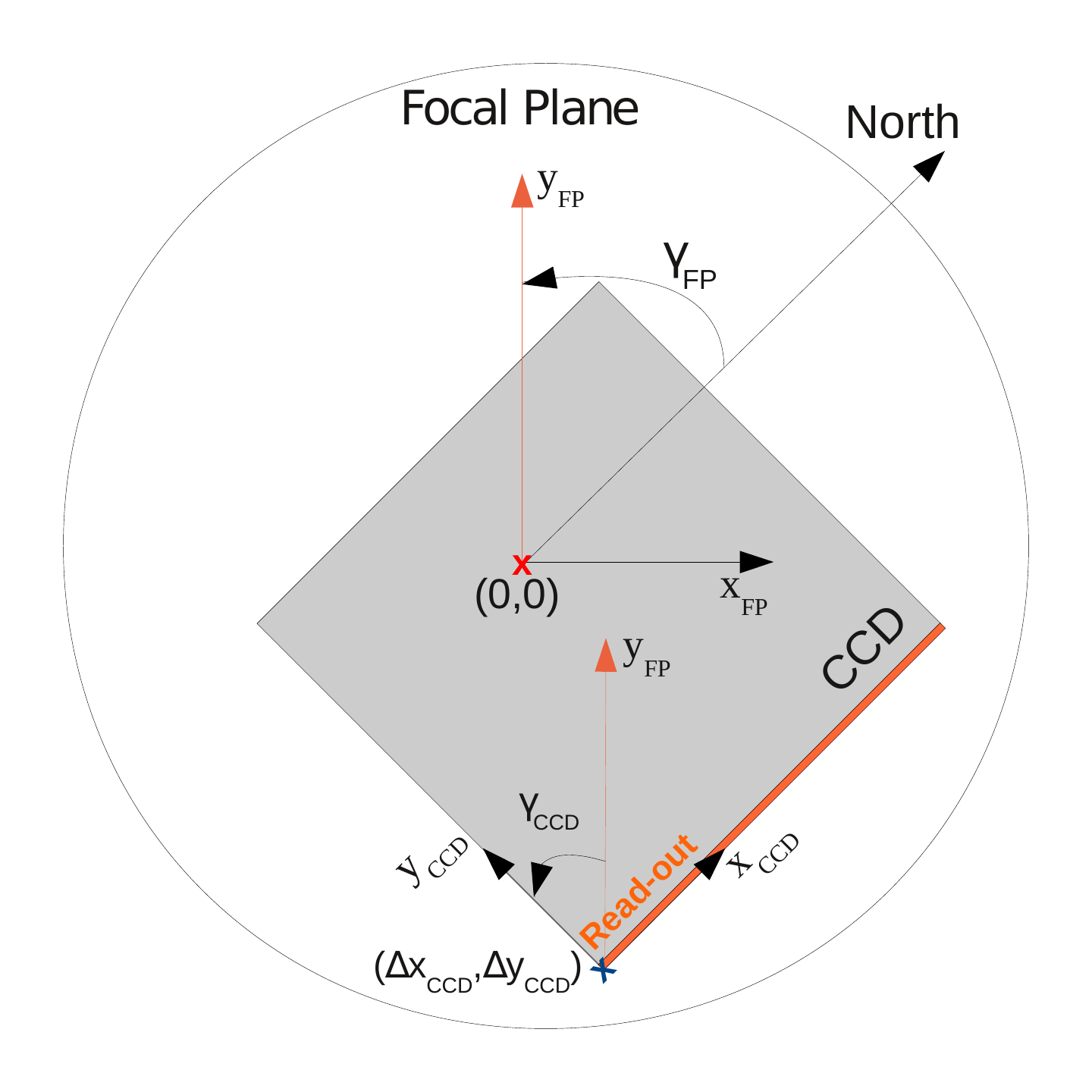}
\caption{Definition of the focal plane orientation and relative CCD location and
  orientation. The focal plane is rotated by the angle $\gamma_{FP}$ with
  respect to the north-direction. The origin of the CCD in the focal plane is
  defined by its offset ($\Delta x_{CCD}$, $\Delta y_{CCD}$) in mm from the
  centre of the focal plane. It is then rotated by the angle $\gamma_{CCD}$
  around its origin.}
\label{fig:focalplane_definition}
\end{figure}

To capture a concrete field, a star catalogue is required as input to time series
project the positions of the stars on the CCD. We used a gnomonic
projection of the sky onto the focal plane, as described below. The orientation
of the optical axis ($\alpha_{OA}$ and $\delta_{OA}$) defines the centre of the
projection. The CCD is not necessarily centred on the optical axis and
  the focal plane has an arbitrary orientation. The position of the origin
(left corner of read-out strip) of a CCD and its orientation in the focal plane
can be arbitrarily defined.  The geometry of this step is depicted in
Fig.\,\ref{fig:focalplane_definition}.

For computational reasons, the full 4510$\times$4510 pixels CCD image
is generally not calculated in the PLATO simulations. Instead, we compute only a
sub-field with a dimension of a few hundred square pixels\footnote{When 128
  sub-pixels are considered, a $100 \times 100$ pixel field contains $(100
  \times 128)^2=163\,840\,000$ sub-pixels, which results in a 1.3 GB memory
  consumption in double precision computations. During convolution with the PSF,
  even more memory has to be allocated.}. This sub-field, as presented in
Fig.\,\ref{fig:fielddef}, is the synthetic image to be written in a FITS
(Flexible Image Transport System) file. To explore different parts of the CCD,
separate simulations with different sub-field coordinates and PSFs have to be
carried out.  Due to the PSF, the brightness of stars close to the
outer margins of the frame affects the pixels close to the inner margins of the
frame. To ensure the inclusion of the flux of these sources, we take an offset margin
around the frame into account.
   
\subsubsection{The CCD sub-pixel matrix}

A CCD consists of a two-dimensional rectangular array of a few million pixels,
which converts the energy released by the photon hits into electron
  counts. Although the pixels typically have a physical size of only a few
$\mu$m square, it is necessary to divide each pixel into sub-pixels during the
simulations to correctly characterize motions smaller than one
pixel. Hence, the intra-pixel sensitivity variations can be approximated by
subdividing each pixel into a number of sub-pixels. The degree of accuracy
increases with the number of sub-pixels. For typical simulations including
jitter pointing variations, we use $128 \times 128$ square sub-pixels to obtain
reliable results. At the end of the image generation, the array is re-binned to
normal pixel space.

\begin{figure}
\centering
\includegraphics*[width=0.45\textwidth, clip,angle=0]{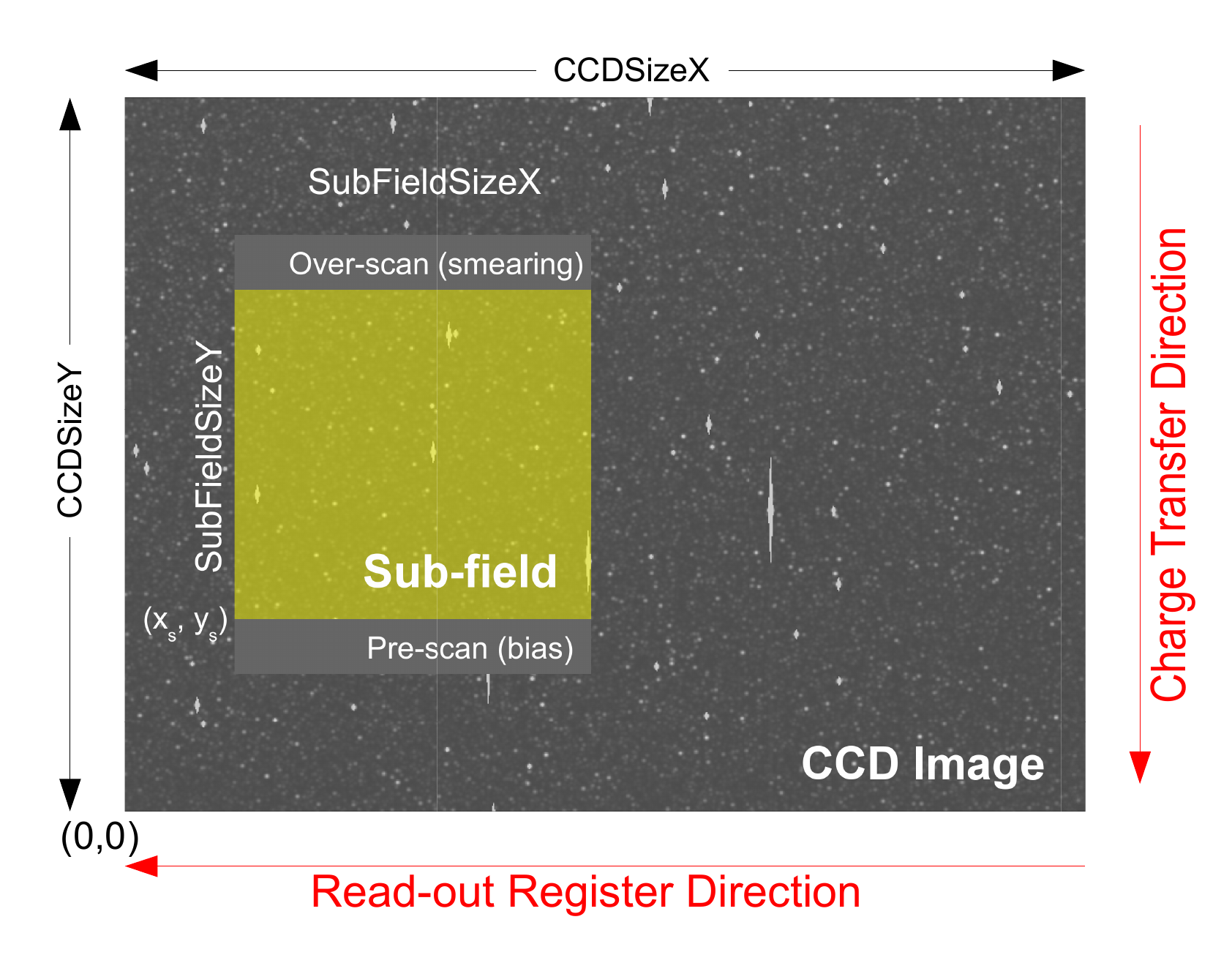}
\caption{Schematic presentation of the sub-field (field of view) and the
  parameters required to define it. The parameters $x_s$ and $y_s$ are the pixel coordinates of
  the origin of the sub-field relative to the CCD.}
\label{fig:fielddef}
\end{figure}

The sub-field is the final image generated; it is a part of the CCD
that is modelled in detail by the {\sc PLATO Simulator} and written to
FITS files (see Fig. \ref{fig:fielddef}).  The read-out smearing effect
  has to be taken into account for frame-transfer CCDs with an electronic
  shutter, as is the case for the PLATO mission. Stars on the CCD that are
outside the sub-field contribute to read-out smearing on the sub-field. Also,
the edge pixels of the sub-field are affected by stars that are just outside the
visible field because of their PSFs. The image convolution with the
  PSF is made for a sub-field that is enlarged by the size of the PSF. Once the
  convolution process has been applied, the sub-field is cut to the size that
  was specified in the parameter file. This also applies for a sub-field that
  lies at the edge of the CCD.

We define the pixel coordinates ($x_i$, $y_i$) in the {\sc PLATO Simulator} with
the following convention: the position of an object in fractional pixels is
defined in a way that integer coordinates lie at the cross section of four
pixels on the CCD (see also Fig.\,4 in the user manual available from the
\href{https://fys.kuleuven.be/ster/Software/PlatoSimulator/user-manual}{\sc
  PLATO
  Simulator}\footnote{\href{https://fys.kuleuven.be/ster/Software/PlatoSimulator/user-manual}{https://fys.kuleuven.be/ster/Software/PlatoSimulator/user-manual}}
web page).

\subsubsection{Mapping stars on the CCD}

Stellar positions are provided through a catalogue that lists their right
ascension ($\alpha$), declination ($\delta$), and magnitude ($m_{\rm v}$). The
sub-pixel position of each star on the CCD is calculated through a gnomonic or
pinhole projection that is widely used in optical astronomy
\citep[see][]{iras1988} and closely reproduces the way in which light is
projected through a lens (or via a mirror) onto a flat surface. We assume that
the optical axis is perpendicular to the focal plane and denote its pointing
direction in right ascension and declination as $\alpha_0$ and $\delta_0$,
respectively. The projection of a point with spherical coordinates $(\alpha_i,
\delta_i)$ onto the focal plane $(x_i, y_i)$ can then be calculated from
\begin{equation}
x_i=\frac{\cos(\delta_i) \sin(\alpha_i-\alpha_0)}{\cos(\delta_0) \cos(\delta_i)  \cos(\alpha_i-\alpha_0) + \sin(\delta_0) \sin(\delta_i)},
\end{equation} 
\begin{equation}
y_i=\frac{\cos(\delta_0) \sin(\delta_i)-\sin(\delta_0) \cos(\delta_i) \cos(\alpha_i-\alpha_0)}{\cos(\delta_0) \cos(\delta_i)  \cos(\alpha_i-\alpha_0) + \sin(\delta_0) \sin(\delta_i)}.
\end{equation} 

With this formalism, the $y$-axis of the focal plane coordinate system is
aligned with the north direction (see Fig.\,\ref{fig:focalplane_definition}). We
then apply a rotation matrix to consider the  arbitrary orientation of
  the focal plane. We consider that the origin of the CCD has an offset from
the optical axis and may be rotated by an angle $\gamma_{\rm CCD}$.  The
position $(x_i, y_i)$ of each star is rounded to its closest sub-pixel
coordinate. This will induce sampling artifacts, that can be reduced with a
higher number of sub-pixels.

For an exposure time of $t_{exp}$ seconds, the flux $F_{\rm phot}$ of each star
is computed from its magnitude $m_\lambda$, the effective light-collecting area
$A$ (in cm$^2$), the transmission efficiency $T$ of the optical system, the
quantum efficiency of the detector $Q$, and the flux per second $F_0$ of a star
with $m_\lambda=0$, from
\begin{equation}
F_{\rm phot}=t_{exp}~F_0~T_\lambda~Q_\lambda~A~10^{-0.4 m_\lambda}.
\label{eq:flux}
\end{equation}  
The value $F_0$ can be determined through numerical integration of the stellar
spectral energy distribution in the relevant wavelength (${\lambda}$) range,
normalized with the flux in that band pass for $m_\lambda$=0.
The quantum efficiency (QE) and transmission efficiency (TE) are the values appropiate  
  for the relevant wavelength (${\lambda}$).

\subsubsection{High-energy particle hits} 
\begin{figure}[!ht]
\centering
\includegraphics[width=0.4\textwidth, trim = 0 80 0 5, clip,angle=0]{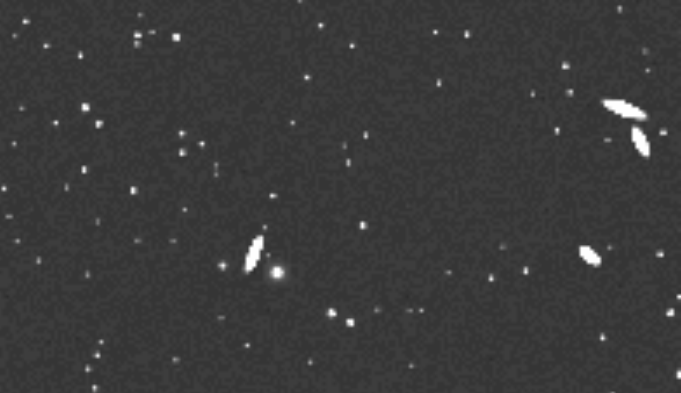}
\caption{Synthesized sample CCD image containing a number of cosmic hits (elongated shapes).}
\label{fig:cosmics}
\end{figure}

High-energy particle hits arising from cosmic rays are a source of noise in
photometric measurements. In space, where no protective atmospheric shielding is
present, cosmic hits are more abundant than on Earth. Estimates for the number
of hits in similar detectors are around two events cm$^{-2}$ s$^{-1}$ for
  CoRoT (see DAK06) and five events cm$^{-2}$ s$^{-1}$ for {\it Kepler}
  \citep{NASA2009}. Such impacts can leave their marks with a large range of
effects and shapes on the detector surface. They can saturate a single pixel or
produce a streak of saturated pixels having complex shapes. We have modelled
proton impacts through a two-dimensional elliptical Gaussian intensity
distribution at sub-pixel level. 
For each event, the major axis of the ellipse,
its central intensity, and two full widths at half maximum values (FWHM) are
determined assuming a stochastic process based on a number of input parameters
to be defined by the user, more particularly 
the frequency of the events and their
intensity. A sample synthesized CCD image containing a number of cosmic hits is
shown in Fig.\,\ref{fig:cosmics}.

\subsection{Satellite jitter}\label{sec:jitter}

Ideally, the orientation of a spacecraft is perfectly stable and does not change
during the monitoring of a stellar field. Unfortunately, this is never the case
and small high-frequency relative pointing variations (jitter) of the spacecraft
cause the images of stars to move on the CCD even during single
exposures. Mainly because a CCD has a non-uniform pixel response,
jitter can lead to a loss of measured flux and aperture photometry can lead to
systematic measurement errors. Fortunately, some methods have been developed to
recover systematic flux variations due to jitter \citep[e.g.,][]{drummond06,
  Fialho2007}, so that it can be corrected for whenever the centroids of the
stars are precisely known.

The jitter is dominated by the reaction wheel noise, structural flexibility, and
star sensor accuracy, whereas the pointing-drift error is mostly due to the
thermal flexibility and variability in the solar aspect angle of the
spacecraft. In order to meet the pointing requirements for PLATO (0.2
  arcsec rms over 14 hours), it is suggested that the reaction wheels are
properly isolated and balanced, and that a model of the thermal deformation is
included in the AOCS (attitude and orbit control system) where the two
  fast cameras will deliver a pointing error signal every 2.5 s.

\begin{figure}[!ht]
\centering
\includegraphics[width=70mm,bb=10 70 595 716, clip,angle=0]{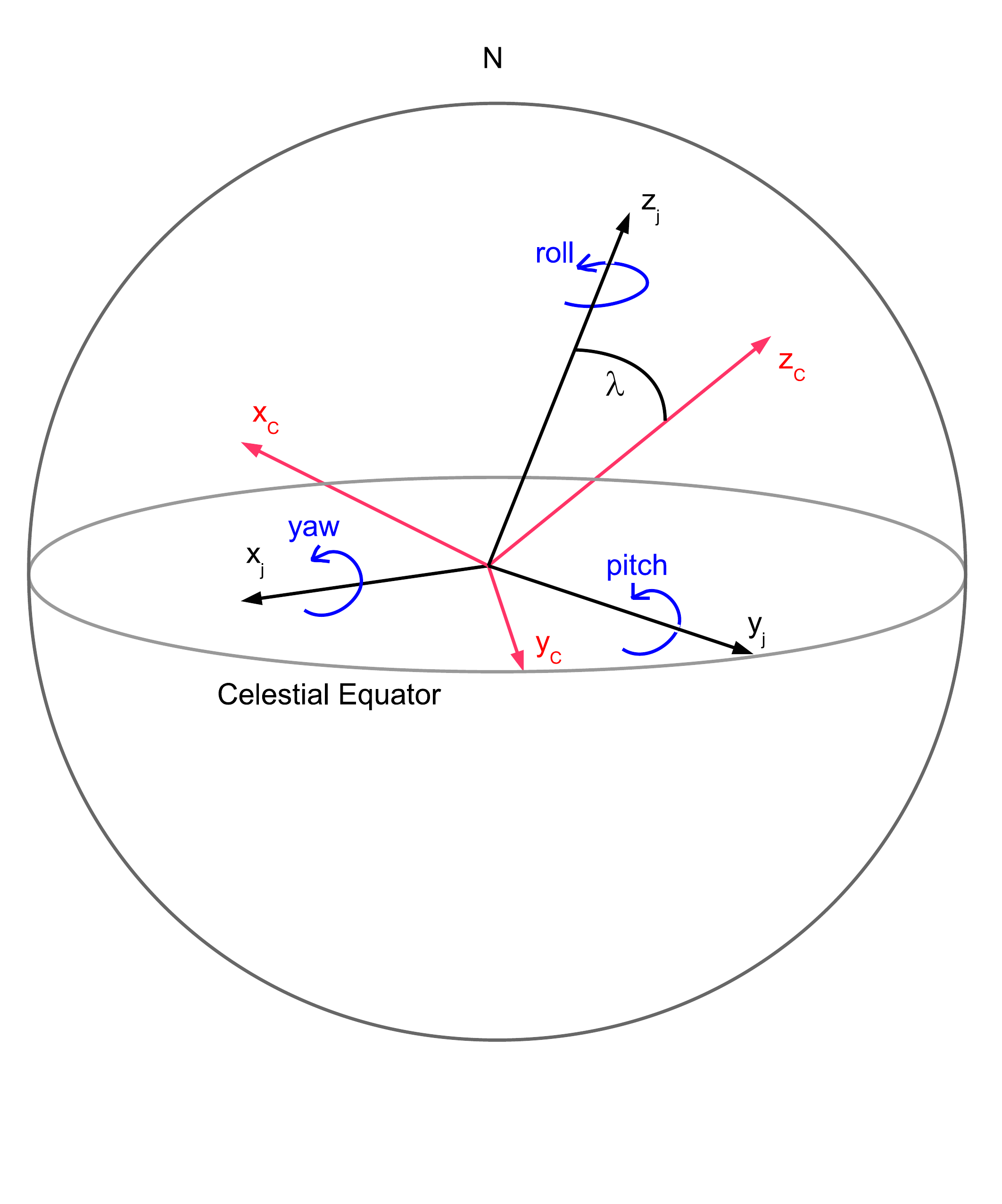}
\caption{Jitter configuration used by the {\sc PLATO Simulator}. The
    jitter roll axis of the spacecraft ${z_j}$ is inclined by an angle
    $\lambda$ to the orientation of the optical axis (${z_C}$). The axis
    ${y_C}$ is normal to ${z_C}$ and lies in the equatorial plane;
    ${x_C}$ is computed from ${y_C} \times {z_C}$.}
\label{fig:jitterdef}
\end{figure}

The effect of jitter on the CCD image of the stellar field is modelled in detail
in the simulator. Jitter movements of the spacecraft can be described by the
displacement angles yaw ($\alpha$), pitch ($\beta$), and roll ($\gamma$). These
angles are defined in such a way that ${z_j}$ points in the direction of
lowest inertia of the spacecraft and is given as angular distance from the
optical axis. The axis ${y_j}$ is normal to ${z_j}$ and lies in
the equatorial plane; ${x_j}$ is computed from ${y_j} \times
{z_j}$. The orientation of the optical axis (i.e., ${z_C}$) is
given in equatorial coordinates and may be different to ${z_j}$. The
${y_C}$ axis lies, like the ${y_j}$ axis, in the equatorial
plane and is normal to ${z_C}$. Finally, ${x_C}$ is computed from
${y_C} \times {z_C}$. The focal plane thus lies in the
${x_C} {y_C}$-plane. Assuming that the telescope is pointed
towards objects that have an infinite distance from the CCD, any rotation of the
spacecraft has the same effect on the focal plane independent of its physical
distance to the jitter axis. Thus, only the angular distance of a star from the
jitter axis is important.

In a next step, the focal plane coordinate system is rotated by the jitter
angles around the jitter coordinate system. First, a rotation around
${x_j}$ with the yaw angle $\alpha$ is carried out. This rotation also
affects the two other jitter axes, ${y_j}$ and ${z_j}$. Next, the
focal plane is rotated around the ${y_j}$-axis (already once rotated) by
the pitch angle $\beta$. This operation also rotates ${z_j}$. Finally, a
rotation around the ${z_j}$-axis (twice rotated) by the roll angle
$\gamma$ is carried out.
 
The position of an object on the CCD is then calculated from a gnomonic
projection on the rotated ${x_C}
{y_C}$-plane. Figure\,\ref{fig:jitterdef} shows a schematic description
of the configuration of the jitter and focal plane reference coordinate
systems. Finally, in order to match the CCD frame of reference,
the $x$ and ${y}$-axes are inverted. Therefore, a
change of yaw moves the field on the CCD in the ${x}$-direction and a
change of pitch moves the field on the CCD in the ${y}$-direction.

In absence of jitter motions, only one image convolution is computed and used
for all subsequent exposures to compute the final image,  including the
  considered noise sources. When jitter is included and modelled, an image
convolution has to be computed for each exposure, evidently leading to much
longer computation times.

\subsection{PSF Convolution}

The {\sc PLATO Simulator} can read a pre-computed PSF from a file or generate a
Gaussian-shape PSF to be used as the PSF mask in the sub-field. Within one
sub-field, all stars are assumed to have the same shaped PSF. This is obviously
an approximation due to different stellar types and because the
shape of the PSF is a function of the position on the CCD.

\begin{figure}[!ht]
\centering
\includegraphics*[width=0.55\textwidth, angle=0]{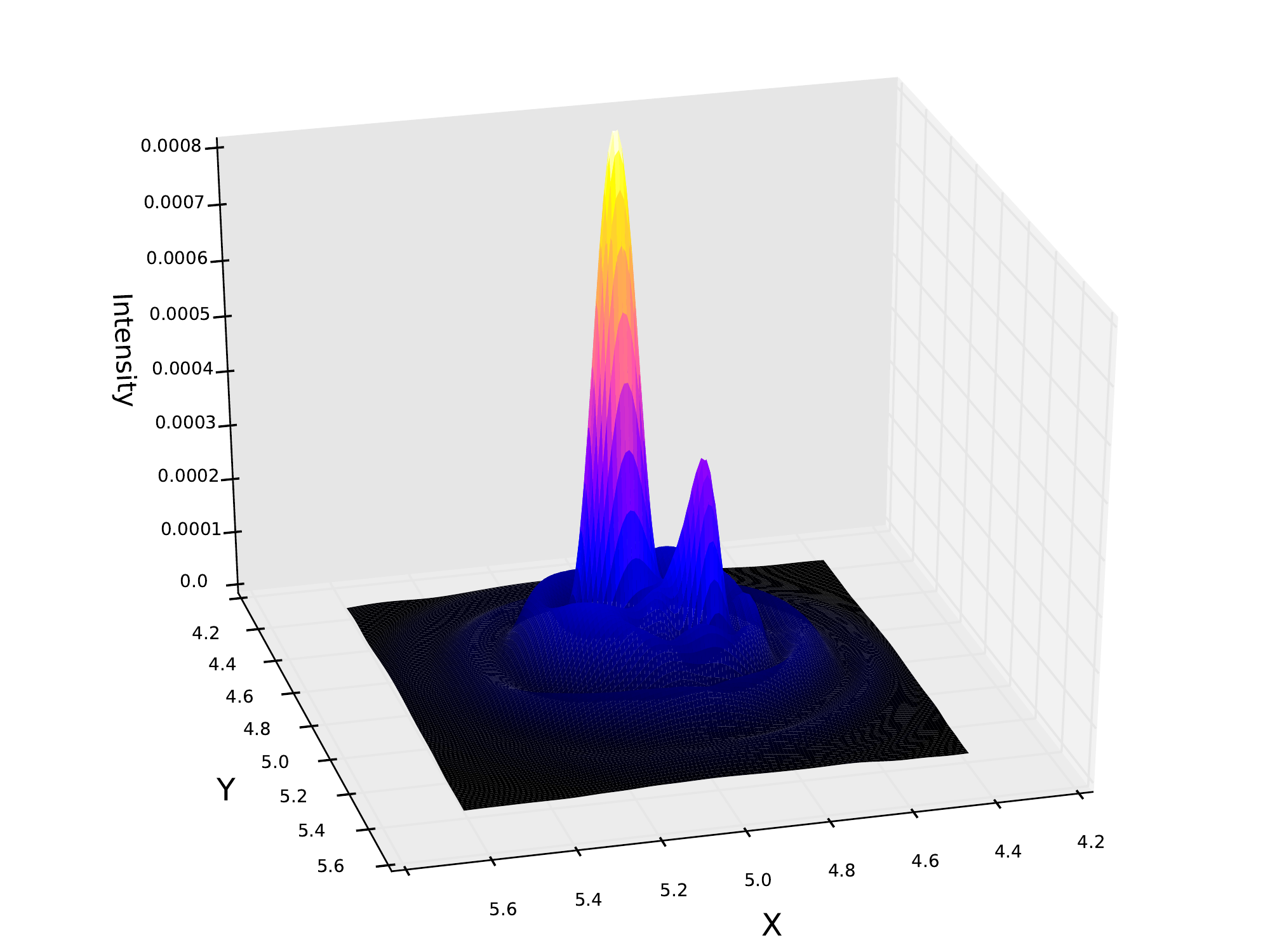}
\caption{Piece of a pre-computed $8\times8$ pixels PSF for PLATO at 2.828 pixels
  distance from the optical axis.}
\label{fig:psf_3D}
\end{figure}

\begin{figure}[!ht]
\centering
\includegraphics*[width=0.3\textwidth, clip, angle=0]{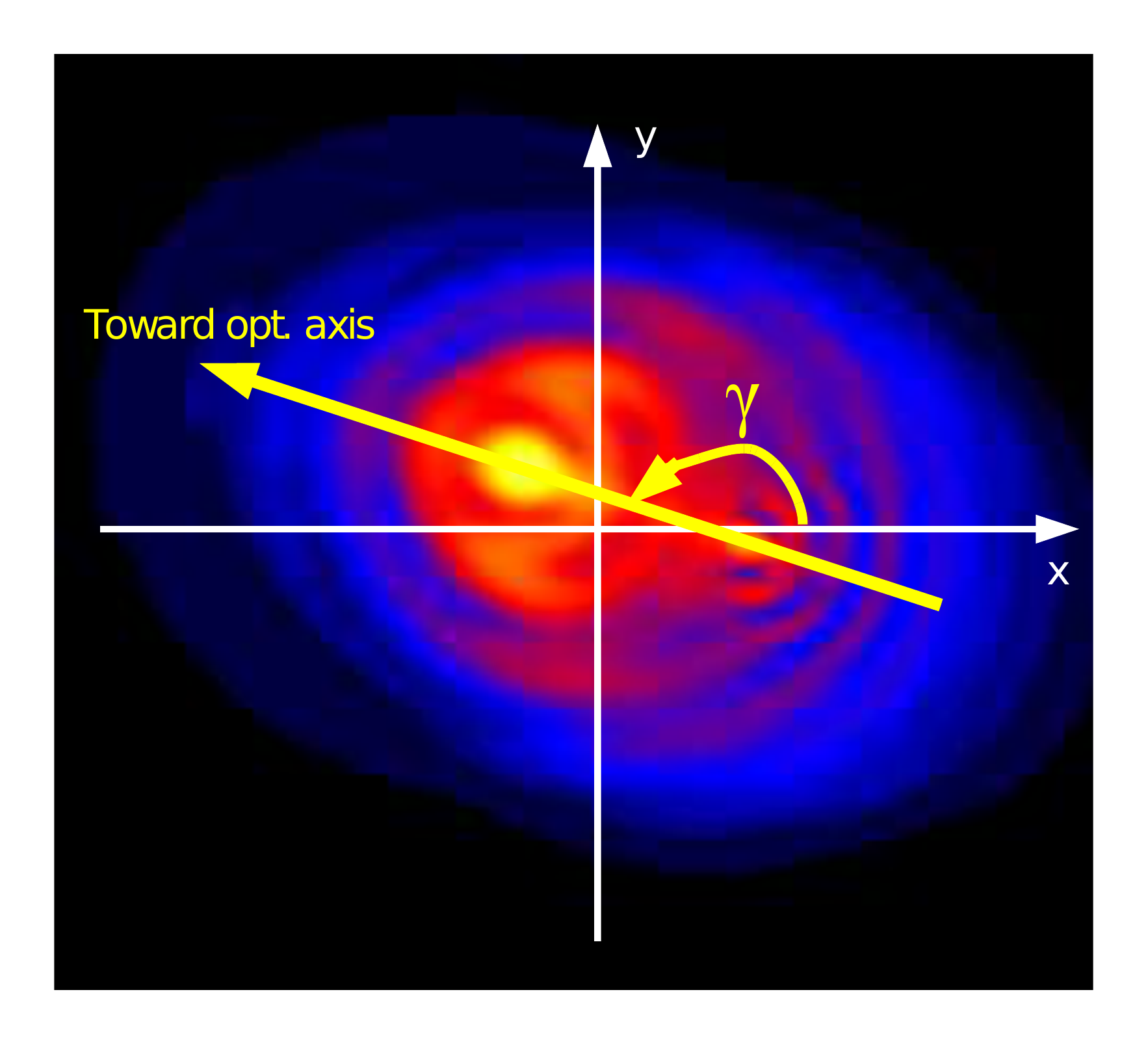}
\caption{Definition of the orientation angle $\gamma$ of
  the PSF towards the optical axis.}
\label{fig:psf_orientation}
\end{figure}

The simulator takes into account that the shape and orientation of the PSF
depends on the location in the focal plane. When a range of PSFs for different
angular distances to the optical axis is provided, the {\sc PLATO Simulator} can
select the PSF that best matches the angular distance of the sub-field centre,
and rotate it in such a way as to correctly orient it relative to the optical
axis (see Figs.~\ref{fig:psf_3D} and \ref{fig:psf_orientation}). Through this
approach, the image distortion that should be included in the PSF pre-computed
mask can be well approximated on the resulting image. It can also shift a
pre-computed PSF by a fraction of a sub-pixel by bi-linear interpolation such
that the centre of the PSF is situated at a cross section. For this procedure,
the centre of the pre-computed PSF has to be indicated. 
Figure~\ref{fig:psf_3D} shows a 3D image of the central piece of a pre-computed
  PSF for PLATO, sampled in a $1024\times1024$ matrix corresponding to
  $8\times8$ pixels (128 subpixels per pixel) at 2.828 distance pixels from the
  optical axis.

For the convolution of the raw stellar field,  the stars are treated as
point sources with sub-pixel CCD coordinates, and the PSF are computed in Fourier
space by applying a 2-dimensional fast Fourier transform (FFT). The
  computation of the product of the field and the PSF in complex Fourier space,
  followed by the conversion of the produced image to real space, is far less CPU
  intensive compared to performing a convolution in real space, particularly
  when a high number of stars occur on the sub-images and the effects of jitter
  have to be used in the simulations.  The drawback is a large memory need, which
  limits the usable dimension of the sub-field and translates into a limit on
  the number of sub-pixels. For practical applications, we kept the product of
  the number of sub-pixels and the side length (in pixels) of the sub-field
  below 6400 (i.e., $50 \times 50$ square pixels and sub-field at $128 \times
  128$ square sub-pixels) for our computations on a dual-core Intel computer
  with 4 GB memory.

As a next step, the sub-pixel matrix, which contains the positions and fluxes of
all stars on the CCD, is convolved with the PSF input mask. The PSF mask should
resemble the real PSF as closely as possible and thus should contain any
artifacts that are due to the optical system. Also, the shape of the PSF depends
on the position in the focal plane, on the wavelength range, and on the stellar
type. For our simulations, a set of mono-chromatic PSFs has been calculated for
different angular distances from the optical axis. From this set, we computed
integrated PSFs for different stellar temperatures assuming a black-body energy
distribution.

The {\sc PLATO Simulator} assumes that the shape of the PSF is the same over the
complete sub-field. The software allows the convolution to be 
performed in real space or in Fourier
space. Real space convolution is carried out by shifting the normalized PSF to
the sub-pixel position of each star and by 
multiplying the sub-pixel intensity by
$F_{phot}$ using the following formalism:
\begin{equation}
F (x_i, y_i)=\sum_{\rm stars}^i F_{\rm phot_{\rm i}} \sum_{x_{\rm PSF}}^j \sum_{y_{\rm PSF}}^k (x_i-x_j, y_i-y_k).
\end{equation}  
The PSF mask is particularly important in the case of on-board photometry
  processing, as will occur for PLATO and was also the case for CoRoT and
  {\it Kepler}. To tackle crowded field photometry processing, PLATO takes a
  6x6 pixel window (or a 9x9 window for the fast cameras) for each target star
  and uses the PSF mask to perform on-board weighted mask photometry. The
  photometry processing is described more in detail in
  Sect.\,\ref{sec:simulations}.

\subsection{CCD sensitivity variations: PRNU}\label{sec:sensitivity}

We adopted the approach described in DAK06 to model the pixel response
non-uniformity (PRNU) of the CCD.  The electronic noise of a typical
  CCD follows a $1/f$ spatial sensitivity distribution. The pixel response
variations across the CCD are typically in the range of a few percent. The
random sensitivity variations of the sub-pixels and a lower intra-pixel
sensitivity for photons that hit the CCD between two pixels has also been taken
into account. The latter two effects can have a significant influence on the
photometry when small sub-pixel displacements due to pointing variations are
present. The PRNU is generally not corrected for in space-borne
astronomy and one must therefore make sure that the pointing of the telescope is
as stable as possible to reduce its impact on photometry.

The {\sc PLATO Simulator} allows us to configure not only the flatfield
  peak-to-peak pixel noise (1.6\% was used in the simulations for PLATO
  discussed below), but also the flatfield sub-pixel white noise and 
   width of the central part of the pixel, which is affected by a loss of sensitivity lower 
   than 5\% compared to pixels away from the edges.

\subsection{Noise effects}\label{sec:noise}
\subsubsection{Read-out smearing}

Frame transfer CCDs that have no shutter are commonly used in space-based
instruments. Because the CCD still receives light during the
read-out, the measured flux of each pixel is increased depending on its distance
to the read-out register. This increase in flux,  the so-called read-out
  smearing ($F_{\rm ROS}$), is proportional to the flux of every pixel in the
  same column but closer to the readout register than the considered pixel
  itself.  Thus, for a pixel in a certain row,
\begin{equation}
F_{\rm ROS}=\frac{\sum_{\rm rows} F_{\rm phot} ~t_{\rm CT}} { t_{\rm exp}},
\end{equation}  
where the summation includes the flux of every pixel ($F_{\rm phot}$) closer to
the readout register than the pixel itself and is proportional to the
charge-transfer time ($t_{\rm CT}$) and inversely proportional to the exposure
time ($ t_{\rm exp}$).

\subsubsection{Sky background}

The brightness of the sky background affects measurements by increasing the
noise of a measurement and setting a lower limit in magnitude at which a target
can be observed with sufficient precision. We refer to \citet{drummond06} for a
description of the background corrections performed for the CoRoT mission.

The sky background (of zodiacal and galactic origin, in units of e$^-$ s$^{-1}$
pixels$^{-1}$) can be set manually or computed from tabular values and
interpolated to the central coordinates of the sub-field. We have adopted the
method by DAK06 for the computations of the zodiacal and galactic
background. The sub-field is assumed to have a constant sky background across
the complete FoV.

\subsubsection{Photon noise}

 Photon or shot noise occurs because of the discrete nature of the electric
  charge carried by the electrons when counting them as representatives of the
  photons that hit the detector, keeping in mind the inherent uncertainty in the
  distribution of the incoming photons. This noise source cannot be avoided. One
  can reduce its effect maximally by collecting a sufficient number of electrons
  at a time, i.e., by increasing the size of the light collecting area or by
  observing sufficiently bright targets. Shot noise follows a Poisson
  distribution and each pixel is treated independent of the other pixels in the
  detector.  Once the theoretical number of photon hits ($n_0$) has been derived
  for a pixel of the CCD (before taking  shot noise into account and after having
  determined all other noise sources mentioned above), this value is replaced by
  a random value taken from a Poisson distribution with mean $n_0$.  The PLATO
  requirement is to reach the photon noise level for stars brighter than
  magnitude 11. This basic requirement, translated into the assessment study of
  the photometric capabilities necessary for the PLATO mission, implies that
  the overall noise level must remain below 34\,ppm per hour for all stars of
  magnitude below 11, as defined in the PLATO Yellow Book \citep{Esa2013}.

\subsubsection{Electronic noise sources}

 Several sources of noise connected with the electronics of the detector
  have been included and modelled in the simulator. These include readout noise,
  full-well saturation, and digital saturation. Their modelling was done in the
  same way as described by DAK06 and can be set to values of choice to perform
  simulations with the {\sc PLATO Simulator} in such a way that they match the
  values determined for the devices used in an optimal way.

An electronic offset (or bias level) of the CCD in terms of analogue-to-digital
units (ADU) is added to the digital signal to avoid negative read-out
values. The electronic offset can be measured in a pre-scan strip that
essentially consists of a few additional read-out rows of the CCD. These rows
only contain the electronic offset and the read-out noise.  A flag can be
  set to zero if the user does not want the pre-scan offset to be added to the
  science frame. The pre-scan strip is defined at the bottom of the sub-field in
  the FITS image that is being modelled in detail (see
  Fig.\,\ref{fig:fielddef}).

 If a pixel receives more electrons than its full-well saturation limit
  (expressed in e$^-$/pixel), we assume the electrons have equal probability of ending
  up in the positive and negative charge transfer direction (termed
  blooming). The electrons reaching the edge of the CCD will not be detected.

The digital saturation limit of the CCD (in ADU/pixel) depends on the
analogue-to-digital (A/D) converter of the detector. For a 16-bit converter, the
digital saturation limit is 65\,536 ADU. The gain of a detector should always be
such that the full-well saturation limit is below the digital saturation limit.


\section{The  {\sc PLATO Simulator} software package}\label{sec:package}

The {\sc PLATO Simulator} is  based upon {\it The Eddington CCD data
    simulator} (\citealt{Arentoft2004} and DAK06), which was originally
  programmed in IDL and was developed for the decommissioned space missions
  Eddington (ESA) and MONS (Danish Space Agency). For realistic applications to
  the PLATO mission, the original software had to be revised appreciably and 
  converted into a much faster computer code. This aspect was
  tackled by \citet{Zima2010}. Moreover, a modern software architecture was
  developed and implemented \citep{Marcos2013}, ensuring  additions, easy adaptability, 
	and use for other missions.

The simulator basically produces a time series of synthetic CCD images based on
the input data for a stellar field, for a telescope, and for instrumental
characteristics, taking  many contributing noise
sources into account. Figure\,\ref{fig:flowdiagram} depicts the flow diagram of the
processing steps to be applied in order to generate a synthetic CCD image of the
stellar field.

Besides the image generation, photometric algorithms were implemented to test
the performance of the simulations and to analyse the created images. A photometry
algorithm measures the flux of each star in the image frame once we correct for the smearing
and background effects.  The delivery of stellar fluxes
  increases the usability and practical value of the simulator given that
  immediate analysis can be performed to assess the quality of light curves.

Although the {\sc PLATO Simulator} has been developed as a multi-mission imaging
simulator, it was constructed in a timely manner to ensure its
usability for the assessment and Phase A/B1 studies of the PLATO
mission\footnote{http://sci.esa.int/plato/}.  In order to accomplish the
multi-mission task, the simulator has been constructed based on  two main
  pilars: the design and use of an architecture based on modularity principles
  and the construction of a common science imaging pipeline.  The modularity
allows the user to treat any of the steps in the processing independently and to add
or modify the implemented functionalities. The design and availability of a
regular common pipeline allows the inexperienced user an intuitive comprehension
of the processing chain and provides easy access to the source code for any
modification or update to be made. For users who want to adapt this simulator
to a particular space mission, it is easy to identify any step in the
process that needs changes or that has different features from those of the present
regular processing. 

Details of the architecture and development of the simulator can be found in
\cite{Marcos2013}.  The current code (written in C++) is run from the command
line. A detailed description on how to use the simulator and the configuration
of the input parameters is given in the {\sc PLATO Simulator} web
page\footnote{https://fys.kuleuven.be/ster/Software/PlatoSimulator/user-manual}.


\section{Applications to the PLATO space mission}\label{sec:application}

The motivation to go to space to acquire photometry of stars for seismic studies
or for detecting exoplanets lies mainly in the lack of atmospheric disturbances
and interruptions due to the diurnal cycle.  A single field in the sky can be
monitored for months to years with a very high duty cycle using the same
instrument, which leads to a homogeneous long-term data set with low photometric
noise levels in the range of $\mu$-mag, while avoiding daily or other alias
structures in the power spectra. Nevertheless, many noise sources remain and
must be quantified through detailed simulations to estimate their impact on the
quality of the observations.

In contrast to the previous high-precision photometers in space, such as MOST,
CoRoT, and {\it Kepler}, PLATO will operate from L2 and will have an
unprecedented large FoV. The main targets of PLATO will be bright dwarfs and
sub-giants with visual magnitudes between 4 and 13 and with spectral types later
than F.  Such bright targets were chosen, not only to facilitate ground-based
follow-up studies, which are essential to confirm the presence of exoplanetary
systems and to pin down the host star fundamental parameters, but also to
construct a database of exoplanet targets bright enough for approved
  future space missions and ground-based facilities to perform infrared
  observations of the exoplanets in transit (e.g., the JWST mission of NASA-ESA
  \citealt{Gardner2006} and the E-ELT project of ESO, \citealt{Snellen2013}).

The goal of the simulations, as presented here, was to predict the quality of
the PLATO space photometry, to assess the transit and stellar variability
detection capability, and to provide essential feedback for the mission
design. Furthermore, the end-to-end simulation can furthermore test the on-board data
processing software and optimize photometric algorithms.

 To illustrate the capacity and value of the simulator, we treat a few of
  the questions that have been raised by the PLATO consortium to test the
  performance of the mission. All the questions below have been tackled with the
  current version of the {\sc PLATO Simulator} through detailed and extended
  simulations. We discuss and place specific emphasis on the last three
  questions in this paper:

\begin{enumerate}
\item How many stars are affected by pollution caused by other sources due to
  confusion during the production of the photometry, as a function of the number
  of stars per square-degree?
\item To what extent does confusion influence the detection of variable stars? 
\item How do different photometric algorithms, e.g., simple aperture photometry
  versus weighted mask photometry, compare with each other?
\item Which optical design performs best?
\item What is the effect of downgrading the number of telescopes?
\item How many stars are observable for PLATO at specific noise levels?
\item What is the effect of jitter on the overall noise budget?
\item What is the variation in noise levels for minor modifications in
  accordance to the prototype detector performance test?
\end{enumerate}

The following section presents the application of the simulator to assess the
answers to the last three questions, i.e., testing the detectability for
different stellar magnitudes, studying the effect of jitter on the noise budget,
and validating the noise levels for the CCD quantum efficiency performances at
different wavelengths.

\subsection{Simulations}\label{sec:simulations}
We conducted a series of simulations to test the performance of the
photometric observations of the PLATO mission in some concrete aspects
regarding the jitter noise,  the influence of the PSF, and the 
CCD performances. 

As an example of one  set of simulations made for the assessment study of
  the mission in terms of photometric performance, we show the results of
  simulations to predict one-week light curves corresponding to 27\,491
  exposures each. We used the photometric algorithms applied to the simulated
  images based on the concrete input conditions of the mission. Some of the
input parameters used in these simulations are given in
Table~\ref{tab:inputparams}.

\begin{table}
\begin{threeparttable}[b]
\caption{Pre-defined properties of the PLATO CCDs in the focal plane.}             
\label{tab:inputparams}      
\centering                          
\begin{tabular}{lc} 
\hline\hline
Input Parameter  & Value  \\
\hline
CCD Size   & $4510 \times 4510$ px \\
Field size & $400 \times 400$ px\\
Pixel resolution\tnote{a} & 1/128 \\
Transmision efficiency & 0.638466 \\
Quantum efficiency\tnote{b}  & 85$\%$ \\
Exposure time & 22 s\\ 
Charge transfer time & 3 s \\
Pixel scale & 15 arcsec\\
Pixel size & 18 $\mu$m\\
Collecting area & 113.09 cm$^2$\\
Flux of $m_\lambda=0$ star & 4\,962\,700 photons s$^{-1}$ cm$^{-1}$\\
Digital saturation & 16\,384 ADU \\
Full well pixel capacity & 1\,243\,000 e$^-$ \\
Gain & 58 e$^-$/ADU\\
Electronic offset & 100 ADU \\
Readout noise & 23 e$^-$ \\
Flatfield pixel-to-pixel noise & 1.6$\%$\\
\hline 
\end{tabular}
\begin{tablenotes}
\item [a] The sub-pixels per pixel parameter has been set to 128.
\item [b] Constant over the entire wavelength range, the value 
was provided by the CCD manufacturer e2V.
\end{tablenotes}
\end{threeparttable}
\end{table}

For these simulations, it was assumed that the CCDs of all telescopes have the
same general properties and that all the noise sources of each of the different
telescopes are the same and occur in an uncorrelated manner.  Since we are
mainly interested in noise estimations of the expected PLATO measurements, the
input stars for the simulations have constant intrinsic magnitudes, i.e.,
  no intrinsic stellar variability was considered. The flux of each star was
  computed through a model that uses its PSF and a weighted mask that gives
strong weight to the central pixels and less weight to the outer pixels. We
derive the expected signal-to-noise ratio from the standard deviation of this
simulated flux.  We evaluate here the performance of the on-board processing
chain, i.e., we considered the data corrected for smearing, offset,
  background, and gain when computing the weighted mask photometry.

The CCD images generated in these simulations have an image size of
$400\times400$ square pixels, corresponding with a field of size $1.67\degr
\times 1.67\degr$. The bottom left corner of this field points towards
$\alpha$=180\degr \,and $\delta$=67\degr. The read-out direction of the CCD is
assumed to be oriented in negative y-direction and the read-out strip is below
the $y$=0 row. The FoV of the CCD determines which stars affect the
sub-field through read-out smearing. The flatfield, pixel response
  non-uniformity (PRNU), is computed by considering a spatial $1/f$-response of
the sensitivity.  For computational reasons, only a small sub-field with a
  side length of 400 pixels was modelled rather than the complete CCD.  The
  jitter of the satellite was derived from recorded time series in yaw, pitch,
  and roll from the CoRoT satellite, transformed to an overall rms of 0.2 arcsec
  and re-sampled at 1 sec intervals to provide sufficient time resolution. 

\begin{figure}[!ht]
\centering
\includegraphics*[width=80mm, trim = 5mm 3mm 15mm 10mm, clip]{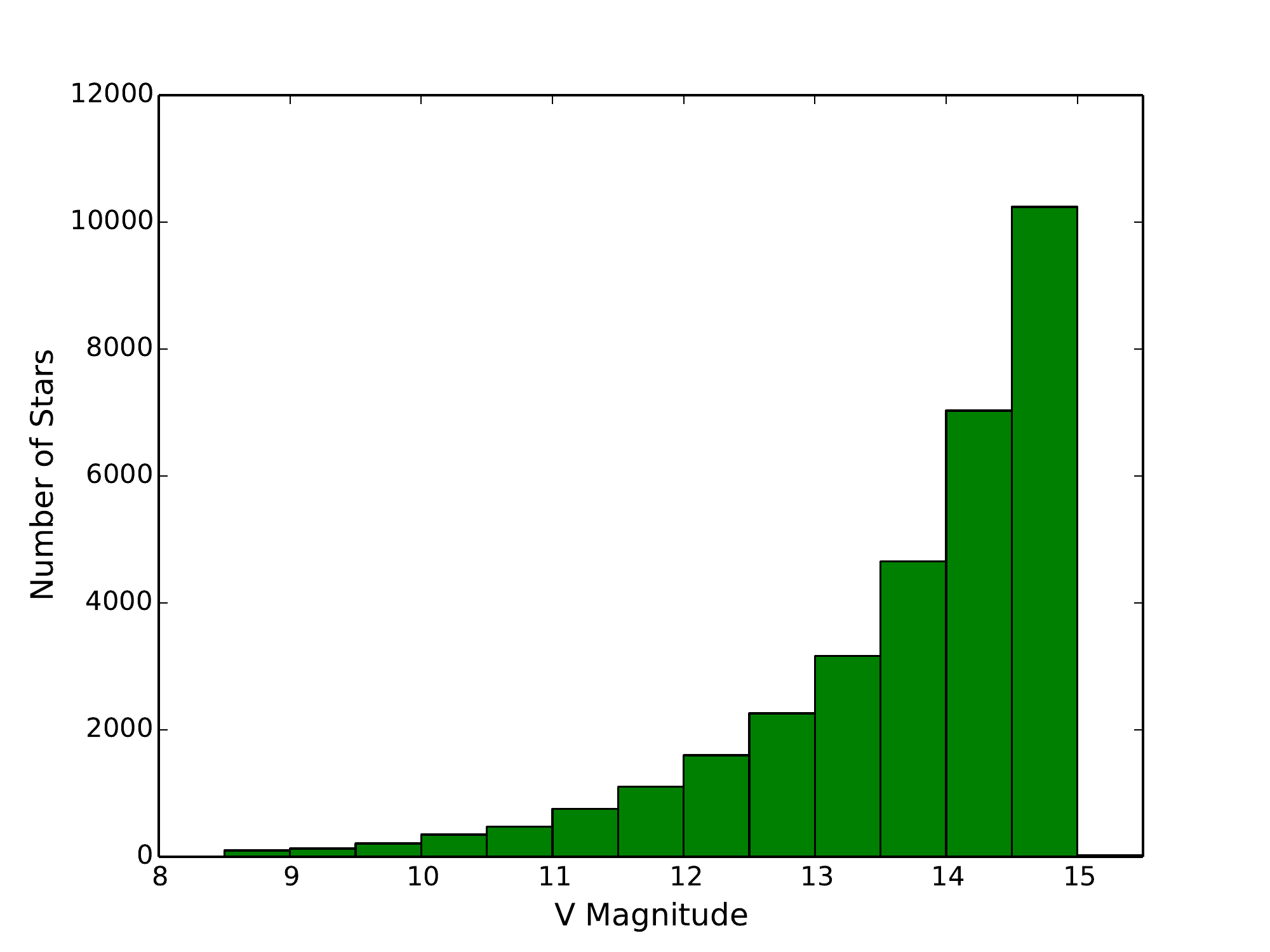}
\caption{Histogram of the stars in the input star catalogue with magnitude above 8
  in bins of 0.5 magnitudes. }
\label{fig:starsmag}
\end{figure}

For these computations, we used a star catalogue of one of the proposed fields of
PLATO containing more than 32\,000 sources with $m_\mathrm{v} \leq 15$ and 1500
stars brighter than magnitude 11. The star catalogue contains their right
ascension ($\alpha$), declination ($\delta$), and magnitude ($m_{\rm v}$); more
details about this field can be found in \citet{Barbieri2004}.  This large set
of stars ensures the computations will provide robust statistics. The number of simulated stars is
given in the histogram in Fig.\,\ref{fig:starsmag} as function of
magnitude. There are only 96 stars brighter than magnitude 8, so these are
invisible in the histogram.

\subsection{Effects of stellar crowding}

Once a simulation of images is performed and each exposure is written in
a different FITS file, the {\sc PLATO Simulator} applies a photometry algorithm
to analyse these generated images.  The flux of each star in the sub-field (see
Fig. \ref{fig:fielddef}) is {\it measured\/} assuming a Gaussian weighted
mask. Subsequently, the noise-to-signal ratio (NSR) and the {\it measured
    magnitude} are derived. The magnitude and photometric flux relation are
  given in Eq.\,\ref{eq:flux}.

\begin{figure}[!ht]
\centering
\includegraphics*[width=95mm, clip,angle=0]{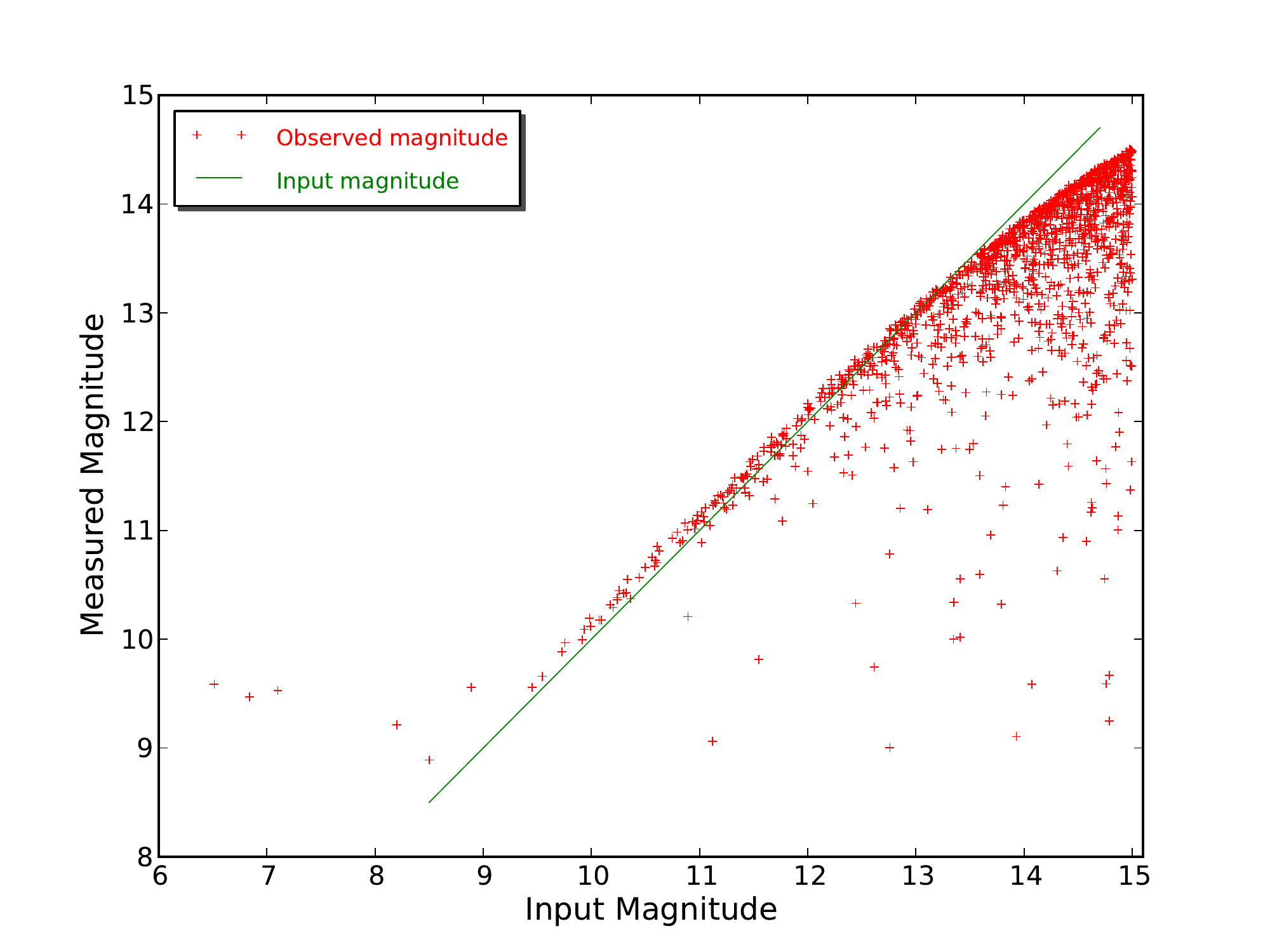}
\caption{Magnitude of the stars in the generated synthetic images as measured
  with the photometric process as a function of the input magnitude given in the
  star catalogue. Each red cross represents the measured magnitude of a star using
  weighted mask photometry. The green line indicates where the measured
  magnitude is equal to the input magnitude.}
\label{fig:measured_input}
\end{figure}

Figure\,\ref{fig:measured_input} presents the magnitude obtained for each source
detected in the output synthetic images as a function of the magnitude of the
same sources in the input star catalogue.  The degradation in performance due to
pollution as a function of the magnitude is represented by the red crosses below
the green line. As the brightness of one star might be affected by another star,
its measured magnitude is lower than its input magnitude.  In the ideal case,
all the stars would lie along the green line.

The sources with input $m_\mathrm{v} \leq 9$ present measured magnitudes above
the input magnitudes due to flux leaking out of the PSF mask as a
  consequence of the smearing effect. The PLATO mission includes two `fast'
telescopes with higher read-out cadence in frame transfer mode to address those
bright sources.

\begin{figure}[!ht]
\centering
\includegraphics*[width=95mm, clip,angle=0]{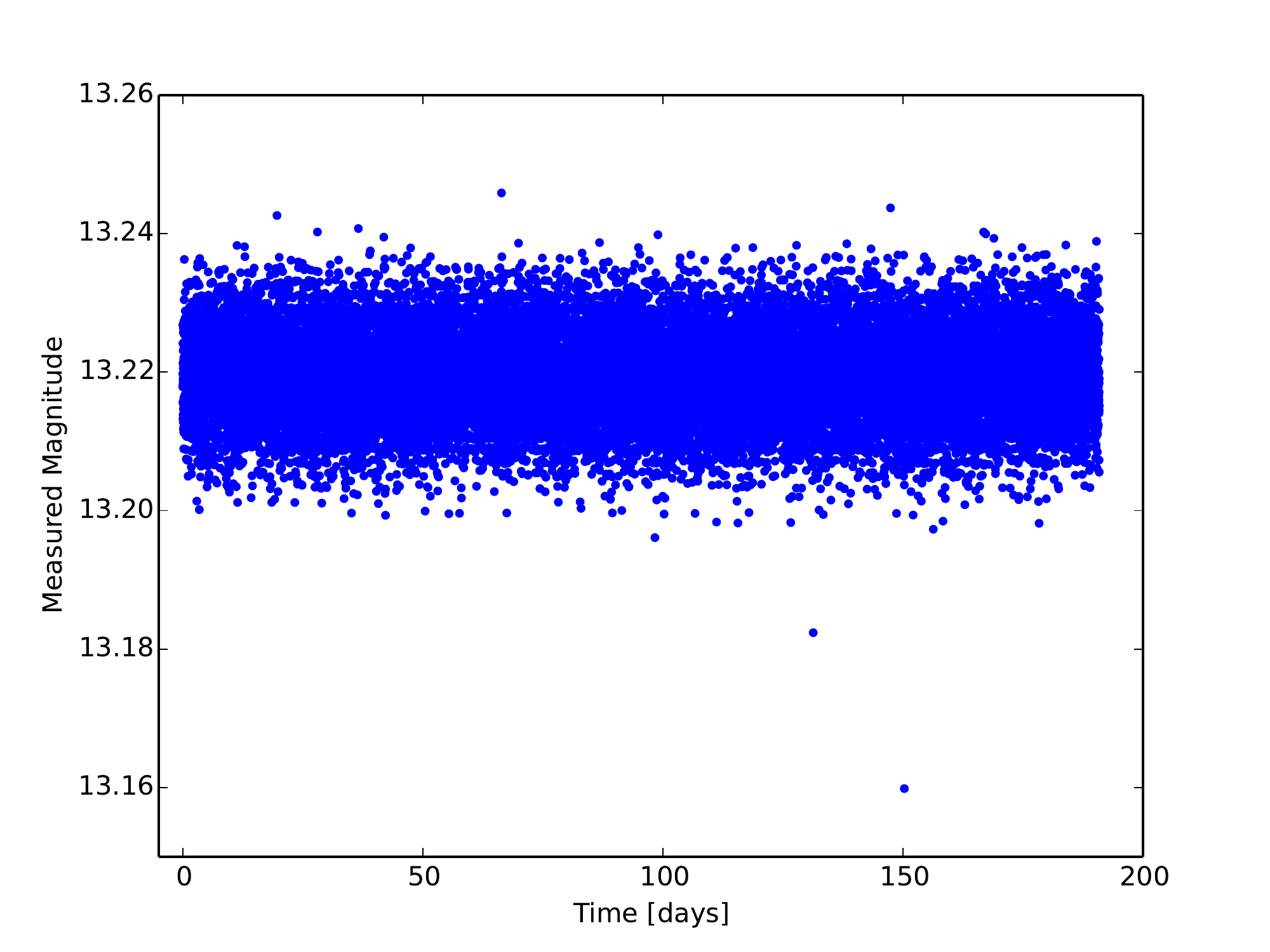}
\caption{Light curve of a 13 magnitude star in the sample. We derive the
  expected NSR of each star from the standard deviation of its simulated {\it
    (measured)} magnitude.}
\label{fig:light_curve}
\end{figure}
\begin{figure}[!ht]
\centering
\includegraphics*[width=95mm, clip,angle=0]{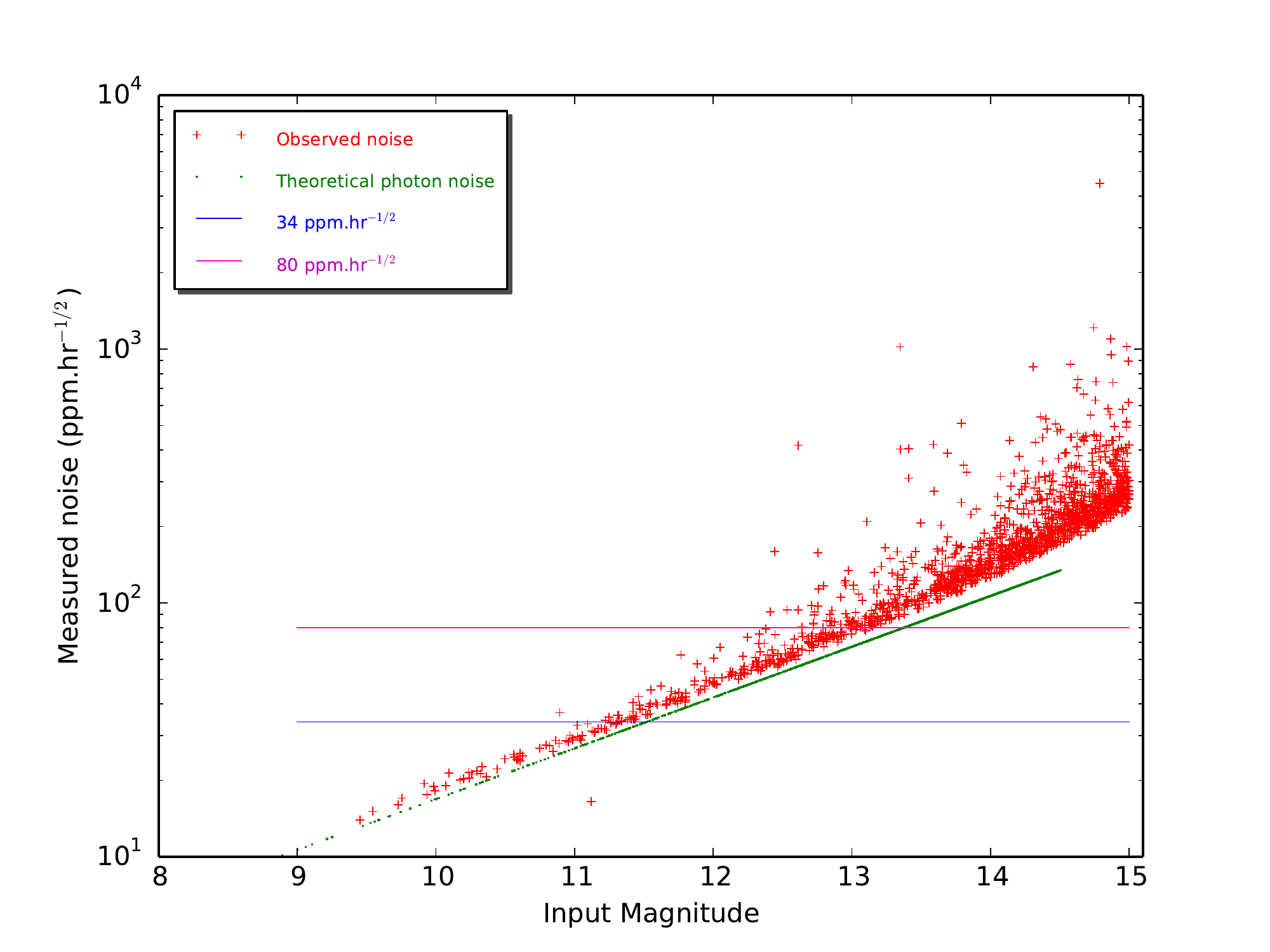}
\caption{Expected noise level in ppm.hr$^{-1/2}$ for observations with 32
  telescopes using noise modeling without jitter and a PSF at the optical
  axis. Each red cross represents the measured noise of a star using aperture
  photometry.}
\label{fig:noise_input}
\end{figure}

For each star, we compute the NSR values as the standard deviation of the
  {\it measured\/} magnitude (see Fig.\,\ref{fig:light_curve}).  The results
are shown in Fig.\,\ref{fig:noise_input}.  The noise has been computed for 32
telescopes, assuming  equal noise properties of the different telescopes and
  their instrument suites and is expressed in ppm.\,hr$^{-1/2}$, given that the
  requirements for PLATO have been determined in ppm per hour of integration and
  the photon noise increases as the square-root of the integration time.  The
red crosses represent the median of the {\it measured\/} NSR of the 27491
exposures for each star in the sample. Green dots represents the expected NSR if
the only noise source was the photon noise.  We also show two
  horizontal lines representing the noise limits at 34 and 80 ppm per hour of
  integration time defined as the mission requirements for the samples with
  $m_\lambda \leq 11$ and $m_\lambda \leq 13$, as defined in the PLATO Yellow
  Book \citep{Esa2013}.  The degradation in performance shown by the increase of
  the measured NSR from the theoretical photon noise limit, due to flux from 
	polluting sources in a crowded field, is as expected.  This
  simulation shows that the requirements for PLATO's priority sample
  ($m_\mathrm{v} \leq 11 $) are fulfilled.

\subsection{Effects of jitter}

We have evaluated the jitter effect that will occur in the science
  instrument from simulations taking  the pointing variations at
  sub-pixel level into account. The jitter behaviour of the spacecraft is described by an
  ASCII input file containing a time series for the yaw, pitch, and roll. The
  adopted jitter time series has been sampled to 1\,s for the cadence in the
  simulations. The pointing error model \citep[see, e.g.,][for more
    explanation]{drummond06} was taken from the ISO \citep{Kessler1996} and
  CoRoT \citep{Auvergne2009} spacecrafts, whose pointing error records are
  available, and rescaled to have an rms of 0.2 arsec in yaw, pitch, and roll
  following the requirements for PLATO.

The noise affecting the images due to the jitter effect has been evaluated by
performing two sets of simulations with the same input parameters, except that
one parameter is with and one is without the jitter option activated. For these
simulations we used a much shorter time series (400 exposures), but maintained
the input conditions as in the previous example, because the larger requirement
of computation time when including the jitter effect. 

\begin{figure}[!ht]
\centering
\includegraphics[width=95mm, clip, angle=0]{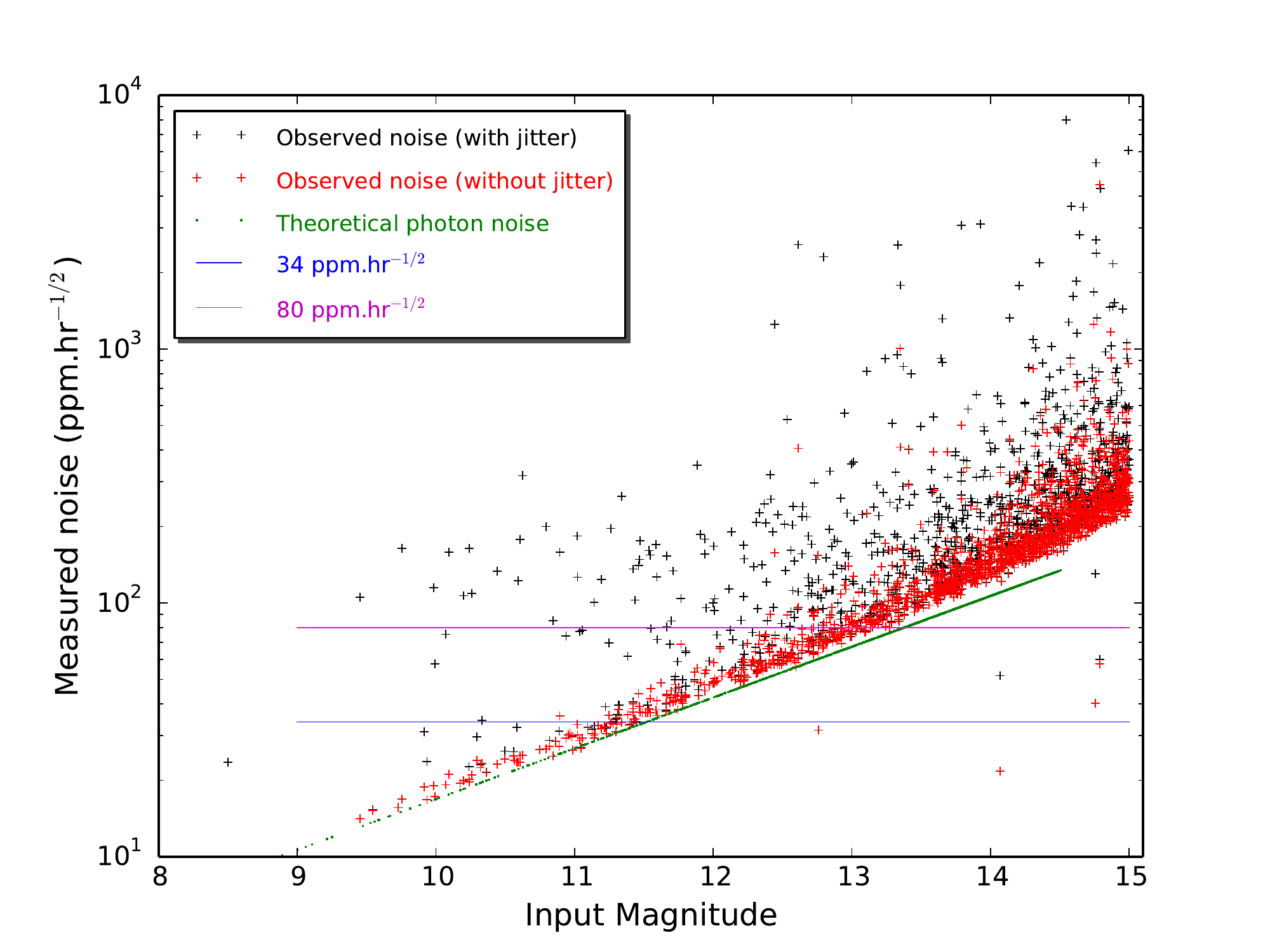}
\caption{Expected noise level in ppm.hr$^{-1/2}$ for observations using noise
  modelling with and without jitter effect for 400 exposures. Each black and red
  cross represent the measured noise of a star using aperture photometry with
  and without jitter effect, respectively.}
\label{fig:jitter}
\end{figure}

 Figure\,\ref{fig:jitter} presents the measured noise for the two sets of
  simulations: one including the jitter noise effect and one without it.  As
  expected, we reach a higher NSR when the jitter effect is present (black
  crosses) compared with the case where no jitter is taken into account (red
  crosses). This is mainly beause of the pointing errors induced by the satellite
  jitter that contribute to the contamination effect and increase the NSR.

We see minor differences when comparing the observed noise without jitter
 in Fig.\,\ref{fig:jitter} (red crosses) with the noise properties shown 
in Fig.\,\ref{fig:noise_input}, which also does not include 
the jitter effect, and was generated from 27491 exposures corresponding
to a one-week time base. These are
due to the more robust statistics from the longer time series, which implies a
better correction.

\subsection{CCDs quantum efficiency variations for different wavelength
  simulations}

 We performed simulations to test the variations in noise levels of three
  different prototypical CCDs that were developed as part of the phase A/B1 of
  PLATO by e2V. For each of those CCDs, a different quantum efficiency response
  at different wavelengths has been provided to us by e2V.  Our intention here
  is not to assess the noise performance of each of the CCDs, but rather to show
  the capabilities of the {\sc PLATO Simulator}. Hence, we maintain the input
  catalogue assumption of constant intrinsic magnitudes ($m_\lambda$) at
  each wavelength ($F_0(\lambda)$ fixed as in Table~\ref{tab:inputparams}) and
  we do not take any intrinsic stellar variability into account.

 Simulations were performed for all the parameters in
  Table~\ref{tab:inputparams}, but taking the quantum efficiency values for each
  of the wavelengths provided to us by e2V (not listed, as we must respect the
  industrial confidentiality of this information).  The noise levels of the
  simulations were evaluated as in the previous examples. 

\begin{figure}[!ht]
\centering
\includegraphics*[width=0.5\textwidth, trim = 15 15 15 5, clip, angle=0]{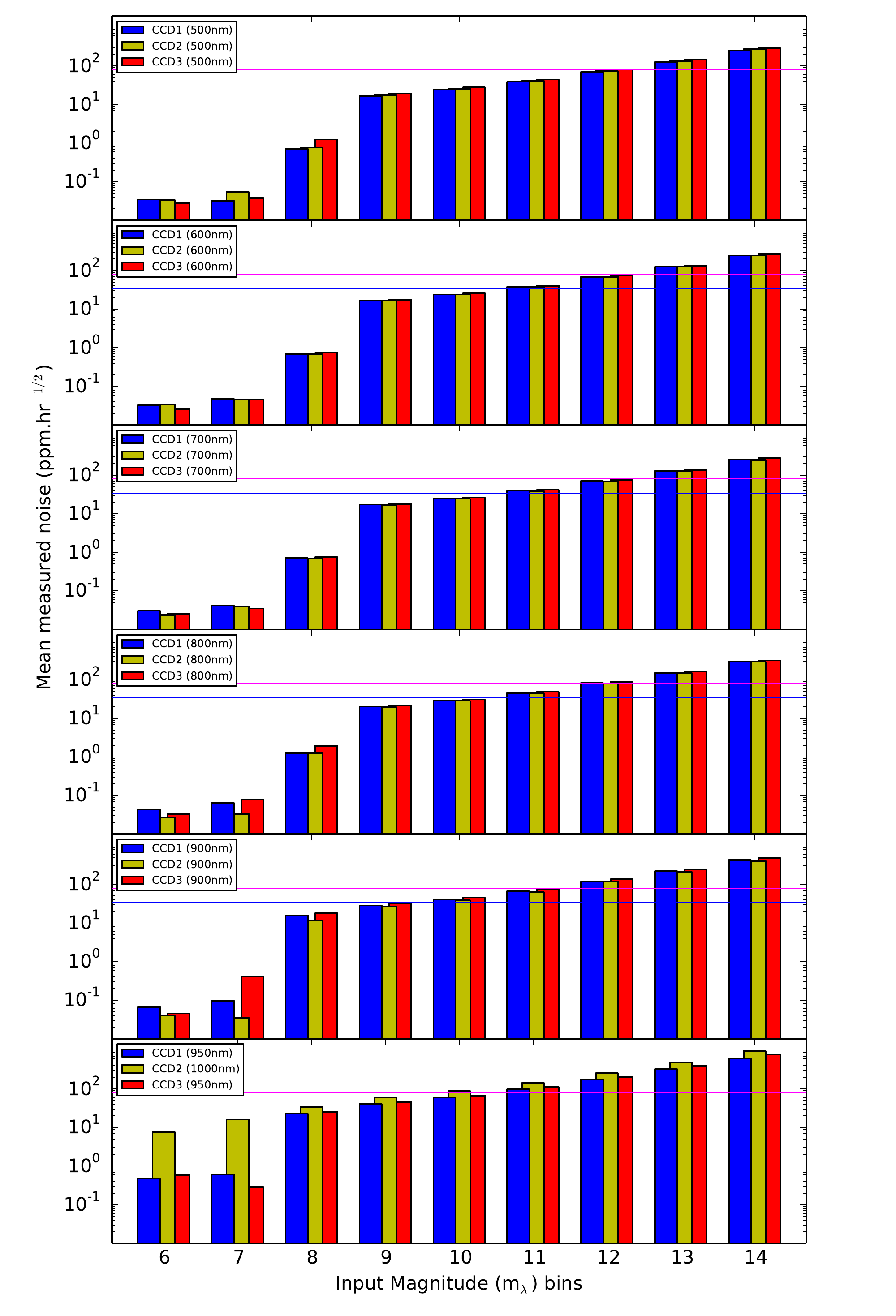}
\caption{Expected mean noise level in ppm.hr$^{-1/2}$ in magnitude bins for
  three CCDs with different quantum efficiencies at different wavelengths. Each
  plot corresponds to a different wavelength and each CCD is represented with
  blue, yellow, or red colour. The two horizontal lines represent the PLATO noise
  requirements at 34 (blue) and 80 (magenta) ppm.hr$^{-1/2}$. }
\label{fig:wavelength_bars}
\end{figure}

In this case, the differences in noise levels are slightly different between
simulations, so we have separated the noise of the stars in magnitude bins and
obtained the mean value for each of those bins to ease the
comparison. Figure\,\ref{fig:wavelength_bars} shows different panels for each of
the simulated wavelengths; each of the plots includes the mean measured noise
for each of the three prototypical CCDs (named CCD1, CCD2, and CCD3). We also
show blue and magenta horizontal lines representing the noise mission
requirement limits for $ m_\lambda \leq 11$ and $m_\lambda \leq 13$ at
34 and 80 ppm.hr$^{-1/2}$, respectively.  The requirements are fulfilled for
shorter wavelengths (see input magnitude bin 12 including stars with $12 \leq
m_\lambda < 13$), but the noise is higher than the
required 80 ppm.hr$^{-1/2}$ for wavelengths above 900 nm because of the decrease in quantum efficiency of the CCDs
decreases.

Fig.\,\ref{fig:wavelength_bars} demonstrates that there is a jump
for $m_\lambda$ from 7 to 8 and from 8 to 9, and that there are further small jumps
from $m_\lambda$=9 onwards. This occurs because these bright sources, which
correspond to sources with input $m_\lambda$ from 6 to 9 shown in
Fig.\,\ref{fig:measured_input}, are affected by the smearing effect, which
implies that flux is leaking out of the photometry mask, providing a higher
measured magnitude. But since the saturation limit is constant, the photometry
deduced from saturated pixels provides a quite stable value and, along with it,
a low noise level. In addition, there are only two, one and three sources for
$m_\lambda$=6, 7 and 8 magnitudes, respectively, which explains the large
dispersion compared to the dispersion obtained when tens or hundreds of sources are
included.

The bottom panel of Fig.\,\ref{fig:wavelength_bars} shows that the
wavelength for CCD2 (in yellow) is 1000\,nm instead of 950\,nm as is the case
for CCD1 and CCD3. The quantum efficiency at 1000\,nm in CCD2 is lower than the
efficiency at 950\,nm for CCD1 and CCD3, leading to an increased level of noise
  represented by the yellow bar in that panel.
 

\section{Conclusions}\label{sec:conclusions}

We have presented the {\sc PLATO Simulator} software package for the simulation
of space-based imaging and photometric analysis with the aim of providing a
versatile tool for the modelling of high-precision space photometry.  The
description of the main noise sources and of the algorithms to transfer these
effects to the synthetic images and generated light curves have been presented
and demonstrated.

We presented some of the results of the application of this tool in the Phase
A/B1 study of the M3 PLATO mission of ESA. Although we only include discussions of the
jitter effect and of the CCD quantum efficiency as
illustrations of the capabilities of the software tool, we used the simulator 
to assess a variety of instrumental and pointing effects to define
the optical design of the mission, its various FoV, the allowable
level of satellite jitter, and the performance of the CCD's electronics and
derived photometry.

The {\sc PLATO Simulator} will be used to carry out future simulations and tests
for the ongoing and upcoming Phase B1/B2 of the PLATO mission project.  The
\href{https://fys.kuleuven.be/ster/Software/PlatoSimulator/}{\sc PLATO
  Simulator} web site includes a detailed description of all the noise effects
and the input parameters to configure those effects, to allow users to perform
new simulations. Installation and user instructions are also included, as well
as the software environment configuration requirements.

We also addressed simulations carried out to evaluate the performance of
  the extension of the original {\it Kepler\/} mission, termed K2. In that work
  (paper in preparation), we paid specific attention to the estimation of the
  expected noise levels due to the pointing stability and possible drift of the
  spacecraft. This additional {\it Kepler\/} study is an illustration of the
  versatility of the {\sc PLATO Simulator} and its ease of use for 
  applications to different space missions.


\begin{acknowledgements}
The research presented here was based on funding from the European Research
Council under the European Community's Seventh Framework Programme
(FP7/2007--2013)/ERC grant agreement n$^\circ$227224 (PROSPERITY) and from the
Belgian federal science policy office Belspo (C2097-PRODEX PLATO Science
Development).  S.S. acknowledges the support of the Belgian National Science
Foundation F.R.I.A.  RS and JG acknowledge the financial support provided by
Centre national d'\'etudes spatiales (CNES) in the framework of the PLATO
project.
\end{acknowledgements}


\bibliographystyle{aa}	    
\bibliography{platosim_aa}		

\end{document}